\documentstyle[aps,twocolumn,epsf]{revtex}  

\def\LAP{\bbox{\nabla}} \def\KA{\kappa^2} \def\Mi#1{{\cal M}^{#1}}
\def\X#1{_{\lower2pt\hbox{$\scrscr#1$}}} \def\Z#1{_{\lower2pt\hbox{$\scr#1$}}}
\def\W#1{^{\raise2pt\hbox{$\scrscr#1$}}} \def\Y#1{^{\raise2pt\hbox{$\scr#1$}}}
\def\ns#1{_{\text{#1}}} \def\Ns#1{\Z{\text{#1}}}
\def\Der#1#2{{#1\hphantom{#2}\over#1#2}}
\def\DDer#1#2#3{{#1^{#3}\hphantom{#2}\over#1#2^{#3}}} \def\JJ{{\cal J}}
\def\NN{{\cal N}}  \def\dd{{\rm d}} \def\e{{\rm e}}
\def\rarr{\rightarrow} \def\GA{\Gamma} \def\PP{{\cal P}} \def\SI{\Sigma}
 \def\goesas{\mathop{\sim}\limits} 
\def\AA{{\cal A}} \def\BB{{\cal B}_\pm(\ph)} \def\UU{{\cal U}(a,\ph)}
\def\CC{{\cal C}} \def\az{\case12a^2} \def\Az{(\az)} 
\def\PNB{\Psi\Ns{NB}} \def\PTL{\Psi\Ns{TL}} \def\PTV{\Psi\Ns{TV}}
\def\Ptv{\Psi_{{}_{\text{TV}}}} \def\PSB{\bar\Psi}
\def\scrscr{\scriptscriptstyle}  
\def\dsp{\displaystyle} \def\scr{\scriptstyle} 
\def\PS{\Psi} \def\pt{\partial} \def\ph{\phi} \def\si{\sigma}
\def\PL#1{Phys.\ Lett.\ {\bf#1}} \def\AP#1#2{Ann.\ Phys.\ (#1) {\bf#2}}
\def\PR#1{Phys.\ Rev.\ {\bf#1}}\def\NP#1{Nucl.\ Phys.\ {\bf#1}}
\def\NC#1{Nuovo Cimento {\bf#1}}\def\CQG#1{Class.\ Quantum Grav.\ {\bf#1}}
\def\const{\hbox{\it const}} \def\psuf{\ph\Ns{suff}}

\begin{document}
\wideabs{
\title{Operator ordering and consistency of the wavefunction
of the Universe}
\author{N. Kontoleon and D.L. Wiltshire\cite{Edlw}}
\address{Department of Physics and Mathematical Physics, University of
Adelaide, Adelaide S.A. 5005, Australia.}
\date{28 July, 1998; ADP-98-43/M71, gr-qc/9807075; {\bf Phys. Rev. D 59 (1999)
063513}} \maketitle \hyphenation{}

\begin{abstract}
We demonstrate in the context of the minisuperspace model consisting of a
closed Friedmann-Robertson-Walker universe coupled to a scalar field that
Vilenkin's tunneling wavefunction can only be consistently defined for
particular choices of operator ordering in the Wheeler-DeWitt equation.
The requirement of regularity of the wavefunction has the particular
consequence that the probability amplitude, which has been used previously
in the literature in discussions of issues such as the prediction
of inflation, is likewise ill-defined for certain choices of operator
ordering with Vilenkin's boundary condition. By contrast, the Hartle-Hawking
no-boundary wavefunction can be consistently defined within these models,
independently of operator ordering. The significance of this result is
discussed within the context of the debate about the predictions of
semiclassical quantum cosmology. In particular, it is argued that inflation
cannot be confidently regarded as a ``prediction'' of the tunneling
wavefunction, for reasons similar to those previously invoked in the case of
the no-boundary wavefunction. A synthesis of the no-boundary and tunneling
approaches is argued for.
\end{abstract}
 }

\narrowtext
\section{Introduction}

Recent developments concerning the possibility of open inflation in quantum
cosmology \cite{HT1,Lin1} have revived an old debate about foundational
issues of that subject \cite{Lin1,HT2,discord}. It is therefore timely to
raise an issue which has been largely overlooked previously, but which in our
opinion has a direct bearing on these foundational issues.

Much of the current debate originates from differing
probability amplitudes calculated in
an approximation which assumes that the transition amplitude for nucleation of
a universe from ``nothing'' is dominated by a Euclidean instanton. Different
probability amplitudes are assumed to correspond to the different boundary
conditions implied by the Hartle-Hawking ``no-boundary'' proposal \cite{HH},
and the ``tunneling'' proposals of Vilenkin \cite{Vil33,Vil37} and Linde
\cite{Lin1,Lin2} which are themselves distinct. In particular, the
nucleation probability for instanton-dominated transitions is assumed to be
\begin{equation}
{\cal P}\propto|\Psi|^2\propto\cases{\e^{-2I\ns{cl}}&$\PNB$\cr
\e^{+2I\ns{cl}}&$\PTL,\  \PTV$\cr}\label{Iproba}
\end{equation}
where the subscripts (NB), (TL) and (TV) refer to the no-boundary wavefunction
and the tunneling wavefunctions of Linde and Vilenkin respectively. For the
solutions in question, which correspond to a model in which gravity is coupled
to a scalar field, $\ph$, with potential, $V(\ph)$ in dimensionless units (see
(\ref{action}) below for our conventions),
the Euclidean action of the instanton is
\begin{equation}
I\ns{cl}={-1\over3V(\ph\Z0)}\,,\label{Iprobb}
\end{equation}
$\ph\Z0$ being the value of the scalar field at nucleation.

According to the standard folklore the above nucleation probabilities can be
determined by a somewhat more careful analysis of the appropriate
minisuperspace models \cite{rev1,rev2,rev3,rev4}. It is this folklore which
we wish to challenge here. We will explicitly demonstrate that the prefactor in
the wavefunction cannot be ignored when calculating the appropriate probability
amplitudes if one is to implement the boundary conditions carefully in
minisuperspace models. In particular, the identification of Eqs.\
(\ref{Iproba}), (\ref{Iprobb})  as representing the relevant probabilities from
which a comparison of the consequences of the competing boundary condition
proposals is to be made, depends crucially on Planck scale physics on account
of such ambiguities. While the ``no-boundary'' proposal turns out to yield a
well-defined probability amplitude independently of such ambiguities,
Vilenkin's boundary condition does not. Vilenkin has previously noted
\cite{Vil33} that $\PTV$ cannot be normalized for one particular
operator-ordering (the ``d'Alembertian ordering''). However, in our view he
appears to have overlooked the full consequences of this issue, which turns out
to quite be a generic problem, as we will show. In particular, it has often
been stated \cite{discord,Vil37,rev4} that operator-ordering is unimportant to
the discussion, especially with regard to probability measures \cite{Vil37}.
Our findings contradict such a viewpoint when it comes to the actual
calculations \cite{Vil37,GrR,Luk,BK1,BK2,bh1} which attempt to
discriminate between consequences of the wavefunction proposals.

Nevertheless, we do believe that some of the viewpoints expressed by each of
the parties to the ``wavefunction debate'' do have some merits. In the last
section of this paper, we will discuss these relative merits in detail, in
light of the mathematical results we will present here. We shall confine our
discussion to minisuperspace, not because we believe that that is the ultimate
arena in which the issue should be decided, but because particular results
which we wish to criticize are derived in this setting, and because even in
more general discussions it is usually semiclassical probability measures
which are nonetheless actually used. Furthermore, while the dangers of too
readily associating $\PTV$ with the semiclassical probability (\ref{Iproba})
have been commented on before \cite{bh2}, it does not seem to have been
appreciated that this can be a problem in even the most well-studied
minisuperspace model, and at the level of affecting commonly claimed
``predictions'' of quantum cosmology such as the prediction of inflation.

\section{Minisuperspace model}

To be more specific, let us consider the 2-dimensional minisuperspace
corresponding to the classical action for gravity coupled to a scalar field,
\begin{eqnarray}
S=&&{1\over4\KA}\left[\;\int\limits_{\Mi{}}\dd^4x\sqrt{-g}
{\cal R}+2\int\limits_{\pt\Mi{}}\dd^3x\sqrt{h}\,{\cal K}\right]\nonumber \\
&&+{3\over\KA}\int\limits_{\Mi{}}\dd^4x\sqrt{-g}\left(
-{1\over2}g^{\mu\nu}\pt_\mu\ph\pt_\nu\ph-{V(\ph)\over2\si^2}\right),
\label{action}\end{eqnarray}
where $\KA=4\pi G=4\pi m^{-2}\ns{Planck}$, $\cal K$ is the trace
of the extrinsic curvature, and the metric is assumed to take the closed
Friedmann-Robertson-Walker form
\begin{equation}
\dd s^2=\si^2\left\{-\NN^2\dd t^2+a^2(t)\dd{\Omega\Z3}^2\right\}\,,\label{FRW}
\end{equation}
where $\dd{\Omega\Z3}^2$ is a round metric on the 3-sphere, and $\si^2=\KA/(6
\pi^2)$.

The Hamiltonian constraint obtained from the $(3+1)$-decomposition of the
field equations may be quantized to yield the Wheeler-DeWitt equation
\begin{equation}
\left[{1\over a^p}\Der\pt a\, a^p\Der\pt a-{1\over a^2}\DDer\pt\ph2
-a^2\UU\right]\PS=0,\label{WdW}
\end{equation}
where
\begin{equation}
\UU\equiv 1-a^2V(\ph), \label{pot}
\end{equation}
and we have allowed for possible operator-ordering ambiguities through the
integer power, $p$, in the first term. The approximation that has been adopted
in previous treatments \cite{Vil33,Vil37} is to confine the discussion to
regions in which the the potential $V(\ph)$ can be approximated by a
cosmological
constant, so that the $\ph$ dependence in (\ref{WdW}) can be effectively
ignored. The resulting equation is then amenable to a standard 1-dimensional
WKB analysis. In this ``de Sitter minisuperspace'' approximation the WKB
solutions are readily found to be
\begin{equation}
\Psi(a,\ph)\simeq{\AA_\pm(\ph)\e^{\mp i\pi/4}\over a^{(p+1)/2}\left[\UU\right]
^{1/4}}\exp\left\{{\pm i\left[-\UU\right]^{3/2}\over3V(\ph)}\right\}
\label{WKBa}
\end{equation}
if $a^2V>1$, and
\begin{equation}
\Psi(a,\ph)\simeq{\BB\over a^{(p+1)/2}\left[\UU\right]^{1/4}}
\exp\left\{{\pm\left[\UU\right]^{3/2}\over3V(\ph)}\right\}
\label{WKBb}
\end{equation}
if $a^2V<1$.

Different boundary conditions will then lead to a solution, $\Psi$,
corresponding to different linear combinations of these WKB components in the
``oscillatory'' and ``tunneling'' regions of the minisuperspace, which
correspond to the oscillatory (\ref{WKBa}) and exponentially dominated
solutions (\ref{WKBb}) respectively.

Both the Hartle-Hawking \cite{HH} and Vilenkin \cite{Vil37} boundary conditions
on the wavefunction require regularity of $\Psi$ as $a\rarr0$. From
(\ref{WdW}) one can see that a potential divergence in the $a^{-2}\Psi,_{\ph
\ph}$ term can be avoided by requiring $\Psi$ to be independent of $\ph$ as
$a\rarr0$. Since $\UU\rarr1$ in this limit, one would thus
na\"{\i}vely expect that the prefactor $\BB$ of
the tunneling WKB modes of (\ref{WKBb}) should take the form
\begin{equation}
\BB\propto\exp\left(\mp1\over3V(\ph)\right)\label{normWKB}
\end{equation}
for the respective modes. There are two problems with this, however. Firstly,
for operator orderings other than the $p\le-1$, the factor $a^{-(p+1)/2}$ in
the prefactor of (\ref{WKBb}) will alter any considerations based on regularity
of $\Psi$. Secondly, the WKB approximation does not hold all the
way down to $a\rarr0$ in any case, and a more careful analysis of the solutions
of (\ref{WdW}) is required in this limit. Such an analysis has been given by
Hawking and Page \cite{HP1} in the case of $\PNB$ with the ``d'Alembertian
operator ordering'' $p=1$, and by Vilenkin \cite{Vil37} for the case of $\PTV$
and $\PNB$ with operator ordering $p=-1$.

We will now extend the analysis of Refs.\ \cite{Vil37,HP1} to both
wavefunctions for arbitrary operator ordering, $p$. In following
\cite{Vil37,HP1} we shall assume that the $\ph$ dependence in (\ref{WdW}) can
be ignored. Such an approximation can be justified if we assume that we are
close to semiclassical solutions (\ref{WKBa}), (\ref{WKBb}) for which $\ph$
varies slowly. This means that the potential $V(\ph)$ should be suitably flat,
which physically is one example of a model leading to ``slow-roll''
inflationary cosmologies. With such a simplification, the $a^{-2}\Psi,_{\ph\ph}
$ term in (\ref{WdW}) is dropped and $V(\ph)$ is approximated by a constant.
It is still not possible to solve (\ref{WdW}) exactly for arbitrary $p$ in
terms of known elementary functions with these approximations \cite{Airy}.
However, it can be solved in a direct fashion in two separate regimes.

Firstly, if $a^2V\ll1$, which for constant finite $V$ will pertain to the
$a\rarr0$ limit, the curvature term dominates the ``potential'' (\ref{pot}),
and with the redefinition
\begin{equation}
\Psi\equiv z^{-(p-1)/4}y(z),\label{ydef}
\end{equation}
where $z\equiv\az$, we find that
eq.\ (\ref{WdW}) reduces to a modified Bessel equation
\begin{equation}
z^2{\dd^2y\over\dd z^2}+z{\dd y\over\dd z}-\left(z^2+\nu^2\right)y=0,
\label{modBessel}
\end{equation}
where $\nu=\pm\case14(p-1)$. The general solution for $y(z)$ is thus a linear
combination of modified Bessel functions, $I_{(p-1)/4}(z)$ and $K_{(p-1)/4}
(z)$.

Secondly, if $a^2V\gg1$, which for constant finite $V$ will pertain to the
large $a$ limit, the $a^2V(\ph)$ term dominates the ``potential'' (\ref{pot}),
and with the redefinition
\begin{equation}
\Psi\equiv x^{-(p-1)/6}w(x),\label{wdef}
\end{equation}
where $x\equiv\case13a^3\sqrt{V}$, we find that
eq.\ (\ref{WdW}) reduces to a Bessel equation
\begin{equation}
x^2{\dd^2w\over\dd x^2}+x{\dd w\over\dd x}+\left(x^2-n^2\right)w=0,
\label{Bessel}
\end{equation}
where $n=\pm\case16(p-1)$. The general
solution for $w(x)$ is thus a linear combination of ordinary Bessel functions,
$J_{(p-1)/6}(x)$ and $Y_{(p-1)/6}(x)$.

The two sets of Bessel function solutions must agree with the respective WKB
solutions (\ref{WKBa}), (\ref{WKBb}) in the limits in which all relevant
approximation mutually hold. Using a combination of an analysis of these limits
and the WKB matching procedure we can constrain the particular linear
combinations of solutions which correspond to the boundary conditions of
$\PNB$, $\PTL$ and $\PTV$.

\section{The ``no boundary'' wavefunction}

Since the
Hartle-Hawking boundary condition \cite{HH} is stated in terms of the path
integral, some further arguments are required to translate this into
boundary conditions on (\ref{WdW}) in minisuperspace. However, the
statement of the relevant boundary conditions on (\ref{WdW}) is uncontroversial
and we will thus follow Hawking and Page \cite{HP1,poetic} in demanding that

(i) in the tunneling region the relevant WKB mode is the $(-)$ solution of
(\ref{WKBb}) only, viz.\
\begin{equation}
\PNB\simeq{{\cal B}_-\over a^{(p+1)/2}\left[1-a^2V\right]^{1/4}}
\exp\left({-\left[1-a^2V\right]^{3/2}\over3V}\right),\label{WKBNBE}
\end{equation}
as is appropriate to the standard Wick rotation $t\rarr-i\tau$ in the
definition of the Euclidean path integral; and

(ii) the wavefunction must be bounded as $a\rarr0$ for all finite values of
$\ph$ and on the past null boundaries of minisuperspace. Thus in a suitable
measure we can take
\begin{equation}
\PNB(a=0,\ph)=1.\label{normNB}
\end{equation}

First consider $p\ge1$. The only modified Bessel function solution of
(\ref{modBessel}) which leads to a regular wavefunction (\ref{ydef}) as
$a\rarr0$ for values of $p\ge1$ is $I_{(p-1)/4}(z)$, yielding a wavefunction
\begin{equation}
\PNB={C_1\over a^{(p-1)/2}}I_{(p-1)/4}\Az\label{smallNB}
\end{equation}
in the $a^2V\ll1$ limit. The constant $C_1$ may be fixed by the
normalization condition (\ref{normNB}) and the small values limit \cite{AbS} of
the Bessel function, giving $C_1=2^{(p-1)/2}/\GA\left(p+3\over4\right)$.

We can now check that (\ref{smallNB}) does agree with (\ref{WKBNBE}) by taking
the limit of both expression for a finite large $a$ for which $a^2V\ll1$ 
nonetheless, which is the limit in which they should match. In practice, this
requires very small values of $V(\ph)\ll1$, i.e., the potential must be much
less than the Planck scale, which is physically reasonable. One finds that
since for finite large $a$ \cite{AbS}
\begin{equation}
I_{(p-1)/4}\Az={1\over\sqrt\pi\,a}\exp\Az\left[1+{\rm O}(a^{-2})\right]
\label{Ilargea}
\end{equation}
the leading term in the appropriate limit of (\ref{smallNB}) does agree with
that from (\ref{WKBNBE}) if
\begin{equation}
{\cal B}_-={C_1\over\sqrt{\pi}}\exp\left(1\over3V(\ph)\right)\,.
\end{equation}

Using the WKB connection formulae 
\cite{Bohm} 
we find
\begin{equation}
\PNB={2C_1\exp\left(1\over3V\right)\cos\left[{1\over3V}(a^2V-1)^{3/2}-
{\pi\over4}\right]\over
\sqrt{\pi}a^{(p+1)/2}(a^2V-1)^{1/4}}\label{WKBNBL}
\end{equation}
for the WKB solution in the oscillatory region, which is the linear
superposition of the modes (\ref{WKBa}) with $\AA_\pm={\cal B}_-$. This can be
checked against
linear combinations of the Bessel function solutions in the limit $a^2V\gg1$.
We find that the solution does indeed match the linear combination of solutions
to (\ref{wdef}), (\ref{Bessel}) given by
\begin{equation}
\PNB={\CC\over a^{(p-1)/2}}\left\{J_{(p-1)/6}(x)+J_{(1-p)/6}(x)\right\},
\label{largeNB}\end{equation}
where $x\equiv\case13a^3\sqrt{V}$ as before, and
$\CC(\ph)\propto\exp\left(1\over3V(\ph)\right)$.

In the case that $p<1$, any arbitrary linear combination of the independent
modified Bessel function solutions $I_{(1-p)/4}(z)$ and $K_{(1-p)/4}(z)$ yields
a convergent wavefunction as $a\rarr0$. Therefore, the Hartle-Hawking condition
does not restrict the wavefunction except by appealing to the semiclassical
behaviour (\ref{WKBNBE}). Since similarly to (\ref{Ilargea}) $K_{-\nu}(z)\equiv
K_\nu(z)$ is given in the limit of finite large $a$ by \cite{AbS}
\begin{equation}
K_{(p-1)/4}\Az={\sqrt\pi\over a}\exp(-\az)\left[1+{\rm O}(a^{-2})\right],
\label{Klargea}\end{equation}
we see that the semiclassical condition only makes the restriction that the
coefficient of $I_{(1-p)/4}(z)$ must be non-zero so as to dominate over
$K_{(1-p)/4}(z)$ in the appropriate limit. If we take the particular
choice
\begin{equation}
y(z)=I_{(1-p)/4}(z)+{2\over\pi}\sin\left((1-p){\pi\over4}\right) K_{(1-p)/4}
\label{WKBy}
\end{equation}
for $p\not\in\left\{-3,-7,-11,\dots\right\}$, then the wavefunction is once
again given by (\ref{smallNB}) as $a\rarr0$ and the previous analysis applies
exactly. For $p\in\left\{-3,-7,-11,\dots\right\}$ the linear combination
(\ref{WKBy}) must be replaced by one for which the coefficient of $K_{(1-p)/4}$
is nonzero, since otherwise we would have $\Psi\rarr0$ as $a\rarr0$, in
violation of (\ref{normNB}). However, provided a linear combination
consistent with the semiclassical behaviour (\ref{WKBNBE}) is chosen, then the
above analysis is not changed is any substantial way. (The exact solutions will
be discussed elsewhere \cite{me}.)

We have thus shown that $\PNB$ can be consistently
defined for arbitrary $p$ in accordance with the approximations usually assumed
for specific operator orderings.

\section{The ``tunneling'' wavefunctions}

Vilenkin's tunneling wavefunction is defined in reverse by placing ``boundary''
conditions in the oscillatory region of the minisuperspace. In accordance
with \cite{Vil37} we require that

(i) in the oscillatory region the relevant WKB mode is the $(-)$ solution of
(\ref{WKBa}) only, viz.\
\begin{equation}
\PTV\simeq{\AA_-\e^{i\pi/4}\over a^{(p+1)/2}\left[a^2V-1\right]^{1/4}}
\exp\left({-i\left[a^2V-1\right]^{3/2}\over3V}\right),\label{WKBTVL}
\end{equation}
so that
${i\over\Ptv}{\pt\Ptv\over\pt a}>0$ there, as required; and

(ii) the wavefunction must be everywhere bounded: 
\begin{equation}
|\PTV|<\infty.
\end{equation}

Beginning with the WKB mode (\ref{WKBTVL}) in the oscillatory region we can use
the WKB matching procedure \cite{Bohm} 
to obtain the appropriate linear combination of the modes (\ref{WKBa}) in the
tunneling region, $a^2V<1$, viz.\ \cite{oops}
\begin{eqnarray}
\PTV=\case12\Psi_-+i\Psi_+,
\label{WKBTVE}
\end{eqnarray}
where
\begin{equation}
\Psi_\pm\equiv{\AA_-\over a^{(p+1)/2}(1-a^2V)^{1/4}}
\exp\left[{\pm1\over3V}(1-a^2V)^{3/2}\right].\label{Adef}
\end{equation}

We can separately match the real and imaginary parts of (\ref{WKBTVE})
with appropriate linear combinations of modified Bessel function solutions to
(\ref{ydef}), (\ref{modBessel}) in the limit that $a^2V\ll1$ with finite large
$a$ using their asymptotic limits (\ref{Ilargea}) and (\ref{Klargea})
similarly to the case of $\PNB$. In this manner, we find that
the appropriate solution in the $a^2V\ll1$ region which corresponds to
Vilenkin's boundary condition is
\begin{eqnarray}
\PTV={\AA_-\over a^{(p-1)/2}}&&\left\{{\sqrt{\pi}\over4}\e^{-1/(3V)}
\left[I_\nu(z)+I_{-\nu}(z)\right]\right.\nonumber\\ &&\qquad\qquad\quad
\left.+{i\over\sqrt{\pi}}\e^{1/(3V)}K_\nu(z)\right\},\label{smallTV}
\end{eqnarray}
where $z\equiv\az$ and $\nu=(p-1)/4$.

The problem with the definition of $\PTV$ is now manifest, since as $a\rarr0$
\cite{AbS}, 
\begin{equation}
K_0\Az\goesas-\ln\Az
\end{equation}
and
\begin{equation}
K_{(p-1)/4}\Az\goesas\case12
\hbox{$\GA\left(|p-1|\over4\right)$}\left(2\over a\right)^{|p-1|/2}
\end{equation}
for $p\ne1$, and so the product $a^{-(p-1)/2}K_{(p-1)/4}\Az$ diverges for
$p\ge1$. In fact, it is quite clear that if we are to have a regular
wavefunction for operator orderings with $p\ge1$ then the only solution to
(\ref{modBessel}) which will yield a regular wavefunction in the limit $a\rarr
0$ is (\ref{smallNB}). That is to say, if regularity of the wavefunction is
important, then any consistent boundary condition for the wavefunction must
coincide with that of Hartle and Hawking \cite{HH} in the context of this
minisuperspace model for $p\ge1$. Any boundary condition which includes a
contribution from the $(+)$ mode of (\ref{WKBb}) in the WKB limit will match
onto the $K_{(p-1)/4}\Az$ solution of (\ref{modBessel}) in the $a^2V\ll1$
limit, and this diverges as $a\rarr0$.
For $p\le0$ the divergence is regulated by the prefactor in (\ref{ydef}),
but for $p\ge1$ the problem is unavoidable. Our conclusion thus applies to
$\PTL$ as well as to $\PTV$.

For operator orderings with $p<1$, (\ref{smallTV}) is well-defined as $a\rarr0$
and thus a normalization condition can be set in this limit to fix $\AA_-(\ph)
$. Vilenkin chose $\PTV\rarr1$ in the $p=-1$ case \cite{Vil37}. However, a
choice $|\PTV|\rarr1$ might be more appropriate here to preserve the real and
imaginary parts of (\ref{smallTV}). In either case, if $V\ll1$, then
\begin{equation}
\AA_-(\ph)\propto\exp\left(-1\over3V(\ph)\right)
\end{equation}
as previously anticipated in (\ref{normWKB}). Only in this manner can the
$\ph$-dependence in the prefactor of the oscillatory WKB wavefunction
(\ref{WKBTVL}) be constrained. The oscillatory
WKB solution (\ref{WKBTVL}) can be matched in the large $a$ limit to solutions of (\ref{wdef}), (\ref{Bessel}) expressed in the combination of a Hankel
function, similarly to (\ref{largeNB}) for $\PNB$.

\section{Probability amplitudes}

\begin{figure}
\centerline{\epsfxsize=7truecm\epsfbox[0 0 283 265]{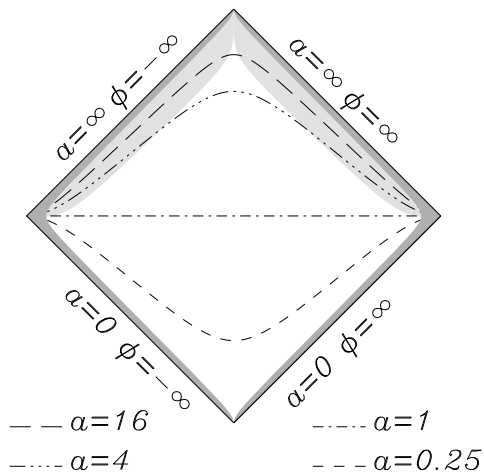}}
\caption{Conformal diagram for $V=0.04\ph^2$. The oscillatory region, given
roughly by
$a^2V>1$, is lightly shaded. Lines $a=\const$ are superimposed. For very large
values of $\ph$ these lie almost entirely in the oscillatory region. The region
of $\ph$-values excluded by a Planck scale cutoff is darkly shaded.}
\label{fig1}
\end{figure}

We now wish to point out that the issue of the regularity of the wavefunction
is crucial in discussions using probability measures in minisuperspace. While
the question of the definition of a suitable probability measure in quantum
cosmology is a tricky one \cite{rev1,rev2,rev3,rev4} it can be argued
\cite{Vil39} that in that in the semiclassical limit the ordinary
``Klein-Gordon'' type conserved probability current
\begin{equation}
\JJ=-\case12 i\left(\PSB\LAP\Psi-\Psi\LAP\PSB\right)\label{current}
\end{equation}
leads to a well-defined probability measure for trajectories peaked around
particular WKB modes, even though $\JJ$ is not positive-definite in general.
The resulting probability density
\begin{equation}
\dd\PP=\JJ\W A\dd\SI\!\X A,\label{WKBprob}
\end{equation}
can be integrated over a hypersurface in minisuperspace to answer statements
of conditional probability such as: ``given that a classical universe
nucleates, what is the probability that it inflates sufficiently ($\sim$ 60--65
efolds)?'' Ideally, the hypersurface $\SI$ here should be chosen in the
oscillatory region, close to the boundary of the tunneling region, but for
potentials satisfying the ``de Sitter minisuperspace approximation'' it is
assumed \cite{Vil37,rev1,rev2,rev3,rev4,GrR,Luk} that this surface can be
approximated by an $a=\const$ hypersurface. (See Fig.\ \ref{fig1}.) In this
limit the probability for sufficient inflation is then assumed to be
\cite{Vil37,rev1,rev2,rev3,rev4,GrR,Luk}
\begin{equation}
\PP\left(\ph\Z0>\psuf\;|\;\ph\Z1<\ph\Z0<\ph\Z2\right)={\dsp\int^{\ph\X2}_{\ph
_{_{\text{suff}}}}\hskip-9pt
\dd\ph\Z0\exp\left(\pm2\over3V(\ph\Z0)\right)
\over\dsp\int^{\ph\X2}_{\ph\X1}\hskip-6pt
\dd\ph\Z0\exp\left(\pm2\over3V(\ph\Z0)\right)},
\label{probinflat}
\end{equation}
where $\ph\Z0$ is the value of $\ph$ at nucleation, $\psuf$ is the minimum
value for sufficient inflation, $\ph\Z1$ is the minimum value for a universe
to nucleate and $\ph\Z2$ a Planck scale cutoff, suggested by the approximations
used. In Fig.\ 1, $\ph\Z1$ and $\ph\Z2$ correspond roughly to the points of
intersection of a suitable $a=\const$ hypersurface with the tunneling (white)
and Planck cutoff (dark) regions respectively.
\vskip-\baselineskip

According to the assumed wisdom the $(+)$ sign in (\ref{probinflat})
corresponds to $\PNB$, and the $(-)$ to $\PTV$, and the resulting probability
is more likely to give $\PP\simeq1$ for $\PTV$ in the presence of a Planck
scale cutoff \cite{Vil37}. This is considered to be a problem for the ``no
boundary'' proposal. However, (\ref{probinflat}) arises from evaluating $\PNB$
and $\PTV$ when peaked around the $(-)$ WKB mode of (\ref{WKBa}) on an
$a=\const$ hypersurface, so that
\begin{equation}
\dd\PP\propto|\Psi|^2\dd\ph\propto{\AA_-(\ph)}^2\dd\ph,
\end{equation}
i.e., in the oscillatory region the phase is unimportant when calculating
$|\Psi|^2$, and it is the prefactor which counts. Our analysis shows,
however, that for $\PTV$ the quantity $\AA_-$ cannot be normalized for operator
orderings $p\ge1$. The problem is thus not merely a mathematical subtlety,
but spells serious problems for the tunneling proposal in terms of its
predictive power.

Of course, it is possible to ``save'' Vilenkin's proposal in
its present form \cite{Vil37,Vil50} if there is some justification as to why
operator orderings with $p<1$ correspond to a natural quantization.
Unfortunately, we know of no such justification. In fact, the only operator
ordering which has ever been claimed to be ``natural'' to date is the
``d'Alembertian ordering'' $p=1$ \cite{HP1,Louko}. Louko \cite{Louko} has made
made a detailed analysis of this point in minisuperspace models, showing that
the ``d'Alembertian ordering'' is preferred if a scale-invarant measure is
chosen when calculating the prefactor by zeta function regularization.

An alternative approach has been pursued by Barvinsky \cite{Bar}, who argues
that the operator ordering question should be determined by demanding
unitarity of the wavefunction. While the issue of unitarity is clearly open to
question in a quantum cosmological setting \cite{Vil39}, it does provide
strong physical grounds on which operator-ordering questions could be
debated. Barvinsky \cite{Bar} has pursued this question in superspace at the
1-loop quantum level. In this context, the ``d'Alembertian ordering'' is
again picked out, this time by the criterion of ensuring Hermiticity of
relevant operators and closure of an appropriate algebra for the 1-loop quantum
constraints.

It is not our intention to focus on the merits of any particular operator
ordering, as any debate must obviously involve questions about Planck scale
physics about which we have, as yet, no direct understanding. However, we
believe that the very fact that a consistent definition of the semiclassical
probability is operator-ordering dependent unless particular boundary
conditions are chosen, does raise some important questions which have been
overlooked in the previous literature.

\section{Wavefunction discord or concord?}

We will now discuss the implications of the result of the previous sections in
terms of the debate about the relative merits of proposals for the boundary
conditions of the wavefunction of the Universe.

Firstly, as mentioned above, Linde's wavefunction, $\PTL$, also suffers from
similar problems to Vilenkin's for operator orderings with $p\ge1$. However, we
consider the criticism about the stability of matter fields in quantum field
theory under a Wick rotation with the ``wrong'' sign, $t\rarr+i\tau$, as
restated most recently by Hawking and Turok \cite{HT2}, as being a much more
serious indictment of Linde's proposal. We will not therefore discuss $\PTL$
further.

There are two levels of criticism which have been put forward by parties to the
debate about $\PNB$ versus $\PTV$. One common criticism of Vilenkin's proposal
is that since its intuition is so closely tied to the WKB approximation in
particular minisuperspace models, it is difficult to suitably generalize it
to superspace. This is due to to the difficulty of rigorously defining the
notions of ``outgoing waves'' and the ``boundary of superspace'' which form
the basis of the tunneling proposal \cite{Vil33,Vil37,Vil50}. Vilenkin has
given arguments to suggest how the tunneling proposal might be put
on a firmer footing, through consideration of the implications of topology
change and other issues \cite{Vil50}. However, the discussion remains
speculative. On the other hand, the no-boundary proposal is not completely
well-defined in a superspace setting either. For example, metrics which are
neither of purely Euclidean nor purely Lorentzian signature must be included
in the path integral to make it converge. Such metrics can make significant
contributions even in relatively simple minisuperspace models, and there is
no obvious unique way in which to define the integration contour through
such saddle points \cite{HalH}.
One must attempt to find a sense in which the Hartle--Hawking proposal can be
reformulated in terms of geometries which are ``approximately'' Euclidean
\cite{Ly}. Using a momentum representation in which the wavefunction depends on
the second fundamental form, as proposed recently by Bousso and Hawking
\cite{BH}, may be a way forward, but much work remains to be done.

It is not our intention to debate the superspace formulation here, as the main
purpose of this paper is to comment on the other level of the wavefunction
debate, which involves the predictions of quantum cosmology. It has become
common in recent papers to simply state that the no-boundary proposal does not
``predict'' sufficient inflation, whereas Vilenkin's tunneling proposal does so
more easily. However, this has not always been the assumed perception, and it
is useful to review how this popular perception arose.

In Hawking and Page's original analysis \cite{HP1} no Planck scale cutoff was
taken in evaluating in the nucleation probability: they set $\ph\Z2=\infty$ in
(\ref{probinflat}), so that the integrals are dominated by the values of $\ph$
above the Planck scale, and $\PP\simeq1$ even for $\PNB$. This argument was
then criticized by Vilenkin \cite{Vil37}, who argued that because Planck-scale
physics goes beyond the semiclassical approximation then a Planck-scale
cutoff must be introduced. Of course, one might still argue, as Page does
\cite{Page} that such a choice is simply an {\it ad hoc} guess about
unknown physics, and the Hawking--Page answer could be the correct one.
However, the use of a Planck scale cutoff for $\PNB$ does seem to be justified
by calculations which suggest that the wavefunction is damped for values of
$\ph$ above the Planck scale by 1-loop effects \cite{BK1,Bar}. The introduction
of a Planck scale cutoff has the consequence that $\PNB$ does not predict
sufficient inflation, at least in terms of the simple models which have been
studied to date \cite{Vil37,GrR,Luk}.

What we wish to stress here, however, is that if one wishes to consistently
exclude predictions based arbitrarily on Planck scale physics from the
discussion, it is not simply good enough to exclude values of $\ph$ above the
Planck cut-off from the $a=const$ integration slice through minisuperspace, one
must also exclude any choices forced by Planck scale physics in the limit $a
\rarr0$.
While it may of course be possible to use conditional probabilities in a way
that avoids the need to normalize the wavefunction \cite{HP1},
the fact remains that the particular chain of argument that
leads to the particular probability measures (\ref{Iproba}), (\ref{Iprobb})
for the minisuperspace model we have studied does rely on the requirement of
normalizing the wavefunction as $a\rarr0$. Thus arbitrary choices about Planck
scale physics via preferred operator orderings enter Vilenkin's proposal as
soon as we require that it make predictions. This point was unfortunately
missed at the time that Vilenkin first discussed the predictions of the
probability of inflation \cite{Vil37} because his analysis at that stage was
restricted to the $p=-1$ model, despite his earlier remark about the $p=1$
case \cite{Vil33}. In Ref.\ \cite{Vil37} Vilenkin stated that since the
Hawking--Page derivation of sufficient inflation from $\PNB$ relied on
contributions from Planck scale energies, the semiclassical approximation on
which the derivation of the no-boundary semiclassical probability density was
based ``could not be trusted in this regime'', and therefore \cite{Vil37}: ``My
conclusion is that at this stage inflation cannot be claimed as one of the
predictions of the Hartle-Hawking approach.'' However, since the consistent
derivation of the semiclassical tunneling probability density also requires
arbitrary choices at the Planck scale, by similar logic we would have to
conclude that at this stage inflation cannot be claimed as one of the
predictions of Vilenkin's approach either. Since the ease of prediction of
sufficient inflation is widely regarded as the principal advantage of $\PTV$
over $\PNB$, we regard this as a rather serious problem for Vilenkin's
proposal.

The strongest claims for the prediction of sufficient inflation from the
tunneling wavefunction have been made from the consideration of 1-loop
effects \cite{BK2}, similar to those leading to the Planck scale cutoff
mentioned above \cite{BK1,Bar}. The claim is that, in the context of a model
with the inflaton non-minimally coupled to gravity, the 1-loop effects lead
not only to a suppression of values of $\ph$ beyond the Planck scale, but also
enhance the bare probability in such a way as to provide a narrow peak in the
probability distribution, thereby leading to sufficient inflation for the
tunneling wavefunction even though the corresponding tree-level probability
does not \cite{BK2}. We believe that our findings place such claims in doubt
for two reasons. Firstly, such calculations \cite{BK1,BK2,Bar} have been
restricted to quantum corrections in $\ph$ about the classical backgrounds with
$\PS\propto\e^{\mp I\ns{cl}}$ and do not address the question of ${\rm O}(\hbar
)$ corrections to $a$ in the limit $a\rarr0$, which were the basis of our
investigation here. Secondly, 1-loop calculations require a choice of operator
ordering: the actual choice of Refs.\ \cite{BK1,BK2,Bar} is the
``d'Alembertian'' ordering, chosen for the requirement of 1-loop unitarity
\cite{Bar} as discussed above, but this choice is at odds with a consistent
definition of the tunneling wavefunction, as we have seen.

The other arena of predictions made from quantum cosmology, which has been the
focus of some debate \cite{bh1,bh2} is the question of primordial black hole
production and the stability of de Sitter space. Our findings here certainly
support the argument of Garriga and Vilenkin \cite{bh2} that $\PTV$ cannot in
general be associated with the probability density (\ref{Iproba}), and thus
criticisms of $\PTV$ based on such a loose association \cite{bh1} are aiming
wide of the target. However, we believe a far better defense of the tunneling
wavefunction would be to find some physical model to which one could
confidently say that $\PTV$ did apply, with definitive predictions. As
discussed above, in our opinion the prediction of inflation does not enjoy such
a status, and we do not know of a physical process which does. While our hopes
for a finding a suitable minisuperspace model for discussing the primordial
black hole issue are more optimistic than the view expressed by Garriga and
Vilenkin, there are many other issues to be considered, such as whether
different horizon volumes are nucleated independently, as these authors have
discussed \cite{bh2}. However, since the relevant discussion of Ref.\
\cite{bh2} again appealed to the probabilities (\ref{Iproba}), (\ref{Iprobb}),
but this time in relation to inflation (which the authors of \cite{bh2}
considered to be justified but which we do not), we believe that many
issues need to be very carefully reconsidered before the debate of Refs.\
\cite{bh1,bh2} could be said to have been put on a firm footing.

Some general comments about the use of probability measures in quantum gravity
are in order. It is common simply to use the bare probability densities
(\ref{Iproba}), often in a saddle-point approximation corresponding to an
instanton, in which both the prefactor and the integration of the probability
density over a hypersurface (or region) of (mini)superspace are neglected.
It is certainly possible to ignore the effects of integration over a
hypersurface if there is a cutoff at a finite scale, such as the Planck scale,
so that the integral is dominated by field values which dominate the
probability density. What is perhaps less well appreciated is that in
considering ``tunneling from nothing'', whether via $\PNB$, $\PTV$ or
otherwise, one is placing a boundary condition at $a\rarr0$ and Planck scale
physics cannot be ignored in this regime. In the discussion of the simple model
here we have seen evidence of this in the important role played by the
prefactor. In more sophisticated treatments there might be other problems.

We consider that the use of instantons as approximations to the calculation of
the amplitude for processes such as pair production of black holes on classical
spacetime backgrounds is well justified since {\it both} the initial and final
states of the system are classical.  However, the nucleation of the Universe is
a different problem in a fundamental sense. To this extent we sympathize with
the sentiment expressed by Linde who likened the semiclassical approach to
quantum cosmology to the problem of the harmonic oscillator, with the comment
\cite{Lin1} that the ``wave function simply describes the probability of
deviations of the harmonic oscillator from its equilibrium. It certainly does
not describe quantum creation of a harmonic oscillator.''

While our findings concerning the prefactor and operator-ordering could be
taken as support for Linde's statement in the absence of a preferred
quantization, we will refrain from suggesting, as
a hard-nosed sceptic might, that the conclusion to be drawn is that
semiclassical quantum cosmology does not predict anything. Rather we believe
that all parties must face up to the fact that boundary conditions at the
beginning of the Universe do entail Planck scale physics by default. In the
case of Vilenkin's proposal this fact is somewhat disguised because the
``boundary'' condition is set in the later Lorentzian regime -- however, as
we have argued, Planck scale physics enters at the moment we wish to make a
prediction. If semiclassical quantum cosmology is to have any pretensions to
make predictions about nucleation of the actual Universe, then boundary
conditions for the wavefunction of the
Universe must be robust when confronted by the Planck scale.
While it remains technically possible that the no-boundary proposal
could suffer from other problems at higher orders in perturbation theory or in
other minisuperspace models, we believe that of the boundary condition
proposals ``on the market'' the prospects for $\PNB$ remain the best, on
account of the fact
that the underlying mathematical intuition in the no-boundary proposal is one
of geometrical smoothness. The ``robustness'' of $\PNB$ vis-a-vis $\PTV$ and
$\PTL$ in the simple minisuperspace we have considered could thus well be
more than an accident.

While the results here seem to have favour the no-boundary wavefunction, or
at least to provide some justification for the use of (\ref{Iproba}),
(\ref{Iprobb}) as the relevant nucleation probability for $\PNB$ in
semiclassical calculations, there are still a number of important outstanding
issues to be resolved in the Hartle-Hawking approach, both on the technical and
interpretational sides. Some of these problems have been mentioned above.
Another major problem is the breakdown of the WKB
approximation, which has been observed to occur in the model with $V(\ph)=m^2
\ph^2$ since the solutions to the Wheeler-DeWitt equation (\ref{WdW}) with $p=1
$ exhibit deterministic chaos \cite{CoS}.

In terms of the question about the semiclassical
probability densities, the most glaring problem which has been glossed over in
the preceding discussion, is the fact that the semiclassical probability
current (\ref{current}) is in fact identically zero for $\PNB$, and to arrive
at (\ref{probinflat}) a decoherence mechanism to the $(-)$ WKB mode of
(\ref{WKBa}) has usually been invoked. If such a mechanism can be found, then
of course the appropriate mode describing the Universe is outgoing in
Vilenkin's sense. The absence of any well-defined mechanism to describe this
decoherence is one of the greatest outstanding problems for cosmological
predictions in the
Hartle-Hawking approach. Since decoherence to a mode that very much resembles
$\PTV$ seems to be what is ultimately desired of the no-boundary approach, one
might hope that a synthesis of the Hartle-Hawking and Vilenkin approaches might
be possible and indeed advantageous. The recent paper of Bousso and Hawking
\cite{BH} could provide a promising start in this direction, because it
suggests a means of distinguishing between the ingoing and outgoing modes of
the wavefunction, thereby suggesting a natural choice of a contour of
integration through complex saddle points in superspace without having to
appeal arbitrarily to decoherence. While we believe the issue of probabilities
in Bousso and Hawking's approach may require more care than they have
exercised, their approach could provide a bridge between $\PNB$ and $\PTV$, and
maybe even eventual concordance.

\smallskip{\bf Acknowledgement} DLW would like to thank J. Louko for his
comments and careful reading of the revised manuscript, and the Australian
Research Council for financial support.


\begin{references}
\bibitem{Airy}An exact general solution can be readily obtained for the special
case $p=-1$, and is expressed as a linear combination of Airy functions
\cite{Vil37}. For arbitrary values of $p$ it is straightforward, although
extremely tedious, to obtain exact series solutions valid for all $a>0$. We
shall present these solutions elsewhere \cite{me}.
\bibitem{poetic}In fact, Hawking and Page \cite{HP1}, who considered the $p=1$
case, only made the assumption of regularity (ii). As we shall show, this is
all that is required if $p\ge1$. For $p<1$, additional assumptions are required
to uniquely specify the wavefunction. We shall therefore choose the
semiclassical behaviour (i) as representing the choice that is generally
assumed to be intended by the Hartle-Hawking proposal.
\bibitem{oops}The expression (\ref{WKBTVE}) differs from the corresponding
expressions given by Vilenkin in Refs \cite{discord,Vil37,rev4,Vil50} for
$p=-1$. This represents a small error in these papers which does not
affect their conclusions (insofar as they only apply to $p=-1$). One may easily
check that (\ref{WKBTVE}) is in fact the correct expression, as it agrees with
the appropriate limit of the $\PTV$ Airy function general solution for $p=-1$
\cite{Vil37}.
\end{references}
\end{document}